# High-Resolution Directional Depth Electrodes: Open-Source FEM Lead-Field Modeling, Characterization, and Validation

Takfarinas Medani, Jace Willis, Chris Wright, Yash Vakilna, Ryan Shores, Raymundo Cassani, Anand Joshi, Richard Leahy, John Seymour, and John Mosher

**Abstract**

**[background]** Depth electrodes used in stereo-electroencephalography (sEEG) and deep brain stimulation (DBS) are essential and evolving tools for neural recording and stimulation. Traditional depth electrodes have limited spatial resolution, typically featuring 8 to 16 cylindrical contacts (0.8–1.0 mm diameter) along a 5–10 cm shaft. This limits recording from small or spatially localized neuronal populations. The recent development of high-density, directional depth electrodes (HD-sEEG) allows improved localization of local field potential (LFP) recordings and spike timing detection. However, understanding their directional sensitivity and validating modeling tools to characterize their lead field (LF) sensitivity remain crucial needs for translation. **[Objective]** We present a comparative analysis of finite element method (FEM) LF modeling for a novel HD-sEEG electrode using two tools: a commercial solver (ANSYS) and an open-source pipeline (Brainstorm-DUNEuro). Our goals are to (i) validate both tools against analytical solutions, (ii) evaluate their differences, and (iii) characterize high-density sEEG directional sensitivity. **[Method]** We modeled LFs of depth electrode contacts in both simple and bio-relevant scenarios. Using Helmholtz reciprocity, LFs were computed by two equivalent methods: (a) electrode-based stimulation with ANSYS, a commercial software, and (b) source-based recording with Brainstorm-DUNEuro, an open-source package well known in the neuroscientific community. We first tested a multi-sphere head model with known analytical solutions to evaluate the two solutions. Next, we simulated an HD-sEEG electrode in a homogeneous volume conductor and computed its LF. Directional gain was assessed by comparing sensitivity with and without the insulating substrate. Additionally, we compared source localization performance using the HD-sEEG electrode against standard sEEG electrodes to evaluate improvements in spatial resolution. **[Results]** In the spherical model, both solvers closely matched the analytic solution, with minimal error and strong concordance. In the realistic model, LF distributions were similar between methods. Modeling the insulating substrate revealed clear directional sensitivity: contacts facing a source had higher sensitivity than those shadowed by the device, demonstrating a "substrate shielding" effect. This effect disappeared when the substrate was conductive, confirming the insulating body's role in modulating signals by source direction. Importantly, the HD-sEEG electrode improved source localization accuracy, as voltage differences across contacts provided reliable directional information. **[Significance/Conclusion]** We show that the open-source Brainstorm platform with the DUNEuro FEM library accurately replicates commercial FEM results for complex LF modeling of depth electrodes, including novel directional arrays. These tools support high-resolution, patient-specific modeling of electrode sensitivity with customizable geometries and insulation, enabling precise analysis of local field potential sensitivity in realistic tissues.

**[Highlights]**

- Comparison of the commercial software Ansys and open-source Brainstorm-Duneuro
- Comparison of a reciprocal and a direct FEM modeling for the sEEG forward approach.
- Lead field solutions computed by both tools are validated using the analytical solution.
- Brainstorm and DUNEuro can compute realistic and high-resolution lead fields at the mesoscale for any arbitrary electrode and insulation geometry
- Validation of the source localization with HD-sEEG and its outperformance compared to standard sEEG models

## 1. Introduction

Advances in neurotechnology are driving the development of multiscale recording and stimulation devices that bridge the gap between microelectrodes and traditional macroelectrodes[1–3]. Two trends motivate improved modeling of these devices' electric fields. First, modern imaging (MRI/CT) provides ~1 mm anatomical resolution, enabling more accurate head models that capture fine details like gray-white matter boundaries and ventricular spaces, which critically influence current flow[1,4,5]. Second, emerging intracranial electrodes now offer *directional sensitivity* and *high contact density*, allowing detection of field potentials on the micro- to mesoscale that was previously unattainable[ref]. For example, the recently developed depth array *di*rectional and *sc*alable microelectrode array (DiSc) has 64–128 contacts arranged radially on a cylindrical shank, permitting recordings from multiple directions around the shaft [3]. This electrode's design, small contacts (<150 μm) on a large insulating substrate, was shown to amplify LFP sources on the side facing a given contact while shielding sources on the opposite side. *Directional LFP recording* capabilities could significantly improve brain-computer interfaces and diagnostics (e.g. epilepsy focus mapping) by highlighting source directionality[3]. Likewise, segmented directional DBS leads have been predicted to provide better localization of neural activity compared to traditional ring contacts[1,6–8].

Finite element modeling (FEM) offers a powerful approach to characterize these complex LFs in detailed head and electrode geometries[6]. However, incorporating sub-millimetric electrode features (insulation, contact geometry) into whole-head models is challenging. Until recently, such modeling often required proprietary software or bespoke code. Here, we leverage **Brainstorm**[9,10], an open-source neuroimaging platform with the **DUNEuro** FEM solver[4,11], and compare it against a gold-standard commercial solver, **ANSYS**[1], for LF computation. Open-source tools like Brainstorm and DUNEuro are attractive due to their integration with neuroimaging workflows and accessibility to researchers without costly licenses. Brainstorm natively handles MRI segmentation, mesh generation, and source analysis, which are non-trivial with general-purpose FEM packages[4,11,12]. By validating Brainstorm/DUNEuro against ANSYS, we aim to ensure that neuroscientists can rely on these free, open-source tools for high-resolution modeling within a dedicated software for the neuroscience community.

We leveraged the *Helmholtz reciprocity principle* as the theoretical bridge between "forward" and "adjoint" approaches to LFs[13,14]. In an EEG context, the *LF* of an electrode describes how a unit current source at any location would be measured by that electrode.

---

[1] Ansys Electronics Desktop (AEDT v.2024 R2)

Reciprocity guarantees this is equivalent to injecting unit current at the electrode and computing the resulting field at the source location. We exploit this equivalence in two independent modeling pipelines to validate both "source recording" and "electrode stimulation" approaches to computing LFs. The validation strategies are informed by similar past validation studies, where numerical EEG forward solutions (BEM, FDM, FEM) were compared to analytical models, often using multi-layer spheres [13,15–21].

In summary, our study addresses three key needs: (1) validating that open-source FEM tools can achieve accuracy on par with commercial software for both conventional and high-density directional electrodes, (2) characterizing the novel directional sensitivity introduced by an electrode's geometry (insulating substrate and dense contacts), and (3) evaluating the practical implications of these modeling results for neural signal recording and source localization. We hypothesize that including the detailed electrode geometry in the model will *significantly affect the LF*, and that both DUNEuro and ANSYS will capture this effect with high agreement. The outcome will inform future *neural engineering* efforts, guiding both device design (e.g., optimal contact layout) and data analysis (e.g., how to interpret LFP amplitude differences across contacts).

## 2. Method

### 2.1. Forward Modeling of Electrical Fields in Brain Tissue

Accurate modeling of electric fields in the brain is essential for both electrophysiological recording (e.g., sEEG, EEG) and electrical stimulation (e.g., TES, sEEG/DBS stimulation). These applications rely on solving the forward problem [5,15,20,22–25], which describes how electrical potentials and fields propagate through complex, heterogeneous head tissues. In this section, we present the mathematical formulations for recording and stimulation scenarios under the quasi-static approximation of Maxwell's equations. We also describe two complementary approaches used to compute the LF based on Helmholtz's reciprocity principle [13,20,26–28].

**i- Electrical Recording Forward Problem (EEG and sEEG)**

The electrical forward problem for EEG and sEEG is derived from the quasi-static approximation of Maxwell's equations, leading to the Poisson equation:

$$-\nabla \cdot (\sigma \nabla u) = -\nabla \cdot j \quad \text{in } \Omega \quad (1)$$

With the following boundary conditions

$$\langle j, n \rangle = 0 \quad on \quad \partial\Omega \quad (2)$$

Here, $u$ is the electric potential (V) $\sigma$ is the conductivity of the medium ($S/m$), $j$ is the current density $Am^{-2}$, $n$ is the unit normal vector at the boundary $\partial\Omega$, and $\Omega$ denotes the volume conductor (i.e., the head).

**ii- Electrical Stimulation Forward Problem (DBS and sEEG Stimulation)**

Electrical stimulation (ES) is modeled using the same quasi-static assumptions [5,20]. When the tissue is considered passive (non-excitable), the intercellular source term $i$ is set to zero, simplifying the Poisson equation to the homogeneous Laplace form:

$$-\nabla \cdot (\sigma \nabla u) = 0 \qquad (3)$$

The electric field is derived from the potential via:

$$E = -\nabla u \qquad (4)$$

In isotropic media, the conductivity σ, is a scalar, while in anisotropic media (e.g., white matter), σ is a symmetric positive-definite tensor that reflects directional conductivity along fiber tracts.

Our goal in both recording and stimulation models is to compute the electric potential distribution throughout the brain volume, either due to internal current sources (recording) or externally applied stimulation (stimulation).

### 2.2. Lead Field Modeling and the Helmholtz Reciprocity Principle

To generate the LF, we employed two complementary FEM modeling approaches to compute electrode LFs, corresponding to the two sides of Helmholtz's reciprocity principle[13,18,26]:

#### i- Dipole-based forward model (recording mode)

A unit dipolar current source is placed in the volume conductor, and the resulting potentials at each electrode contact are calculated (simulating an EEG recording). Repeating this for three orthogonal dipole orientations yields the LF vectors at that source location.

#### ii- Electrode-based adjoint model (stimulation mode)

A unit current is injected at a given electrode contact (with a return path at a reference electrode), and the electric field vector is sampled throughout the volume conductor. This directly produces the LF as a vector field ratio to the applied current. In this manner, the value of the LF at any point is proportional to the electric field (E-field) there when 1 A is applied at the electrode.

Both approaches should yield identical LF vectors for each source location, providing a powerful validation. In practice, we implemented the dipole-based model in Brainstorm/DUNEuro (which natively supports dipole forward solutions) and the electrode-based model in ANSYS (which easily simulates current injection).

*Figure 1* illustrates this reciprocity; the same vector field can represent either the current flow from a stimulating electrode pair or the sensitivity fields of that electrode pair to a current source.

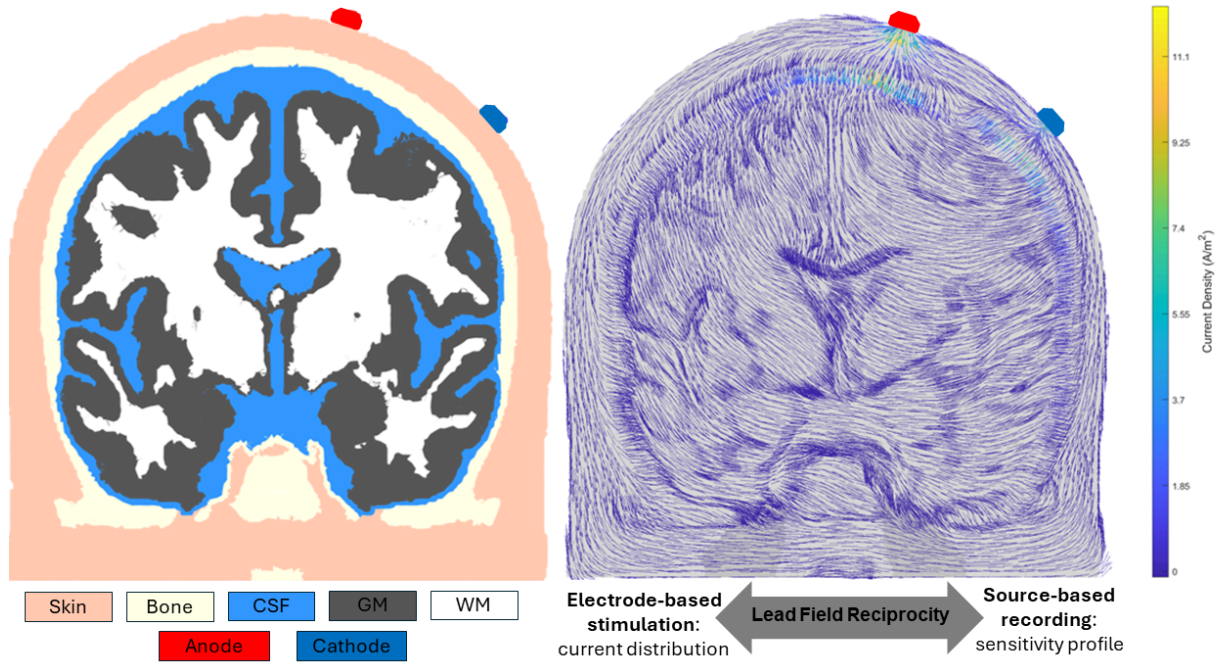

**Figure 1**. *Left*: Realistic head model generated for FEM simulation with five tissues, namely, from inner to outer: wm, gm, csf, bone and skin. *Right*: illustration of the Helmholtz reciprocity theorem. The same vector field represents, on the one hand, the current distribution induced by two stimulation electrodes (red patch: anode, blue patch: cathode) and, on the other hand, the sensitivity profile of these two electrodes to sources at any given location. This figure was generated using the Electrical Simulation by plotting the electrical field computed using the roast toolbox[29,30].

e illustrate the Helmholtz reciprocity theorem (Helmholtz 1853) in the context of electrode stimulation and electrode recording. We consider a simple stimulation scenario where currents are injected and discharged via two electrode patches over the right (anodal red) and left (cathodal blue) temporal brain areas. The figure shows the calculated distribution of current intensity and current orientation in the brain, computed in a realistic five-compartment (skin, skull (bone), CSF, Gray matter, and white matter) head model. Importantly, the same distribution has an equally valid interpretation representing the sensitivity profile or the LF of an EEG recording with only the two electrodes placed in the same positions. More precisely, the vector in a certain voxel (described by position, orientation, and magnitude/length) illustrates which potential difference between the two electrodes would be measured if an assumed neuronal source with a unit strength of 1 were oriented in the direction of the LF vector. For a current of any orientation, the projection onto this LF vector can be computed to yield the measured potential. For instance, a current dipole oriented perpendicular to the LF vector orientation would generate a measurable difference in potential between the two electrodes (projection equals zero).

### 2.3. Numerical Simulation and Validation with Spherical Head Models

For the initial validation, we employed a multi-layer spherical head model, where quasi-analytical solutions to the forward problem are available[24,31,32]. Two configurations were tested: a single homogeneous sphere and a concentric 4-layer sphere representing the scalp, skull, cerebrospinal fluid (CSF), and brain. Radii and conductivities were chosen according to standard values[15,17,33], with the brain layer having an outer radius of 78 mm and conductivity of 0.33 S/m, the CSF layer at 80 mm and 1.79 S/m, the skull at 86 mm and 0.01 S/m, and the skin at 92 mm with conductivity 0.43 S/m. For the single-layer case, an identical conductivity of 0.33S/m was assigned across all layers. A total of 200 electrodes were uniformly distributed on the outer surface of the sphere. We defined a set of test dipole sources at varying eccentricities within the volume (Figure 2 and Figure 8). EEG forward potentials were computed using two methods:

### i- Brainstorm/DUNEuro

We created a standard brainstorm protocol and created and used the spheres as head models for the forward computation. The FEM mesh is generated using Brainstorm and the iso2mesh plugin. Dipoles of unit moment were placed at each source location within the inner layer, then the leadfield is computed using the DUNEuro FEM solver with the standard FEM parameters. The resulting potentials were computed at all 200 electrodes. In Brainstorm, the solver directly returned the LF matrix (channels × (sources *3 orientations)).

### ii- ANSYS

The same spherical model and electrode configuration were implemented. One ampere (1 Ampere) of current was injected at each electrode in turn, using a distributed reference across the sphere surface to approximate an infinite reference. The resulting electric field and potential were recorded. From these, LF vectors were derived based on the electric field response to each stimulation pattern.

For both EEG and ES configurations, analytical solutions for spherical models [Ref] served as ground truth. The numerical potentials from each solver were compared against these analytical values for selected dipole positions and orientations.

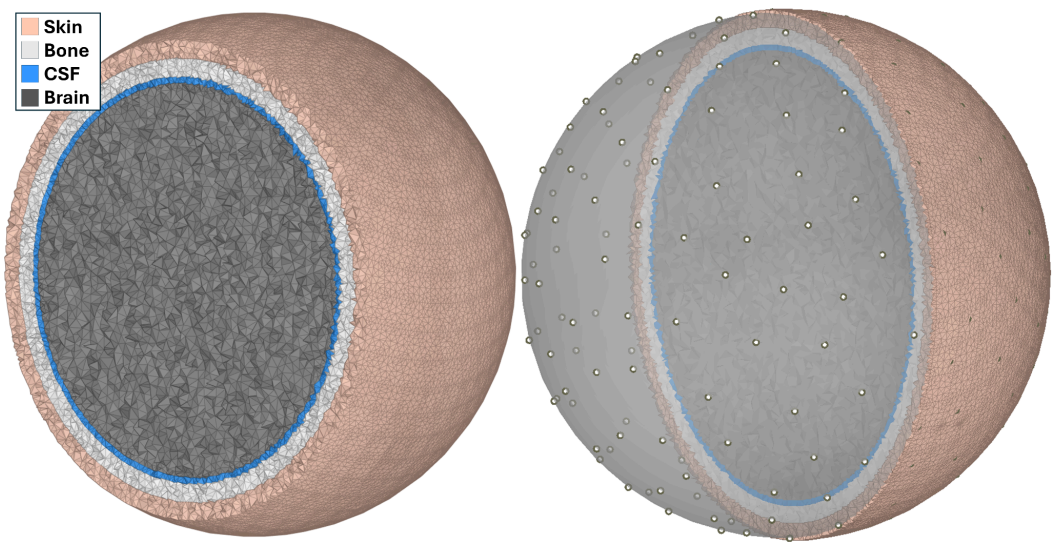

**Figure 2.** *Left*: Spherical Head Model setup with 4-layer in Brainstorm representing the main tissues, from inner to outer: brain, csf, bone, and skin, with comparison to analytical. *Right:* the sphere model with 200 electrodes on the surface.

### iii- Evaluation Metrics:

Quantitative agreement between solutions was evaluated using three standard metrics: (1) the Relative Difference Measure (RDM), which quantifies angular differences between normalized LF vectors (ideal value = 0); (2) the Magnitude Error (MAG), which captures discrepancies in field magnitude (ideal value = 0); and (3) the Correlation Coefficient (CC), reflecting overall similarity between LF patterns. High concordance across these metrics in this idealized setup validates the solvers and establishes a baseline for testing in more realistic head geometries.

### 2.4. Numerical Simulation of the High-Resolution sEEG

Next, we modeled the HD-sEEG to investigate LF differences in a realistic scenario. The electrode model was based on a design referred to as a *di*rectional and *sc*alable (HD-sEEG) array, reported in recent literature[3]. It consists of a cylindrical polyimide substrate (diameter 0.8 mm, length 10 mm) with 128 electrode contacts (0.12 mm diameter pads) distributed in four columns and 32 rows along the shaft (with every other row staggered about 45 degrees to form an 8-column diamond configuration). The contacts on one side of the cylinder are thus physically oriented 180° opposite those on the other side. We used both Brainstorm[9] and Iso2mesh[34] for the creation and visualization of these models in Brainstorm, and DUNEuro FEM solver for the leadfield computation. Figure 3 shows the electrode model: a red cylinder (electrode shaft) inserted into a cubic volume of brain tissue (gray) that we will use as the region of interest (ROI), with yellow representing contacts on its surface. We meshed the volume with tetrahedral elements of ~0.5–1.0 mm edge length. The insulating substrate was assigned a very low conductivity (~1e-8 S/m), and the surrounding medium was set to brain conductivity. A similar model was created and simulated within Ansys[35].

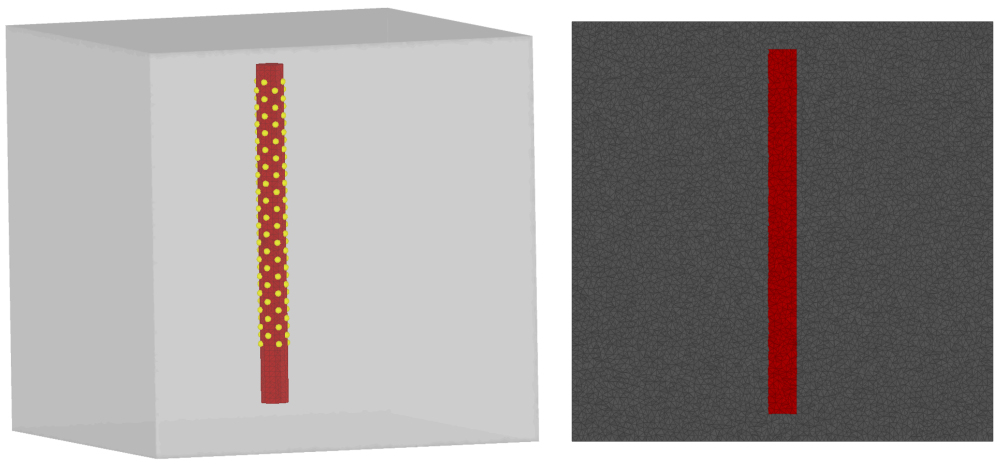

**Figure 3.** Visualization of the 3D HD-sEEG electrode model in Brainstorm and the bounding box representing the region of interest (ROI) in the brain where the device is simulated. ***Left***: 3D perspective view of the model. The contact electrodes are represented by the yellow spot around the body of the electrode in red. The HD-sEEG device(in red) is inserted into the brain tissues (gray colored). The ROI of the brain here is modeled by a cubic shape with 12mm on each side. ***Right***: Lateral side with perpendicular view to the electrode and the ROI. We generated a regular mesh resolution for both the ROI and the HD-sEEG. The source space (location of the dipoles) is formed by a regular grid spaced by 0.25mm within the ROI as shown in Figure 4. Finally, the bounding box layer is set to be the reference plane to simulate an infinite reference in our FEM modeling.

Similarly to the spherical model, we computed the LF of this model using the two software packages as follows:

### i- ANSYS Simulation (electrode-based)

We sequentially applied a 1 A influx current source to each contact on the HD-sEEG with a return point on the boundary and computed the electric field throughout the volume. Due to the linearity of Maxwell's equations in steady-state, each simulation yields one column of the LF matrix, the sensitivity vector field for that contact. We repeated the simulation for all contacts to build the full LF. By reciprocity, this maps the voltage response to a unit dipole at all measured points in space, which is considered the point-wise gain vector. The LF is then formed by sampling the electric field vector at all the source points in the volume conductor (see green dots as shown in Figure 4). The exported electric field vector is processed with a ratio of currents, where multiplying by any source and dividing by the simulation current recovers a gain vector.

### ii- Brainstorm/DUNEuro Simulation (source-based)

Test dipoles of unit strength were placed on a 3D grid of source points within the volume (green dots in Figure 4). From Brainstorm and using DUNEuro-FEM routines with a continuous Galerkin formulation and the Venant source model, potentials at all contacts were computed for dipoles oriented along the x, y, and z axes. The same outer boundary reference as in ANSYS was used to ensure consistency as the "infinite" far-field reference in the FEM modeling. This yielded LF vectors for each source location.

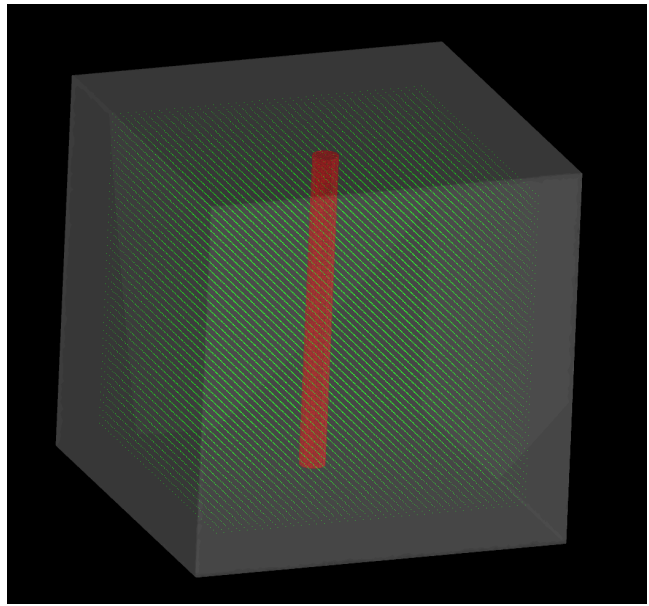

**Figure 4.** Visualization of the Brainstorm/DUNEuro source-based FEM model with the ROI, the source space and the HD-sEEG. 68921 test dipoles of unit strength (green dots) are placed on a 0.25mm 3D regular grid of source points within the brain volume (transparent cube) referred to as the region of interest. The cube's outer boundary (blue) serves as the "infinite" far-field reference to ensure consistency with the ANSYS model. The red cylindrical shape represents the directional electrode array embedded within the volume. This setup is used to compute LF vectors for dipoles oriented along the x, y, and z axes at each source location.

## 2.5. Simulation Study: Dipole Source Localization Using Dual HD-sEEG

In this section, using Brainstorm, we systematically investigated dipole source localization using simulated dipolar sources activation recorded by two HD-sEEG arrays, contrasted with conventional sEEG configurations. The two HD-sEEG arrays, each containing multiple recording contacts, were positioned at a fixed spacing of 6 mm apart within a finite element method (FEM) volume conductor (ROI) created in Brainstorm (see figure 5). We defined eight dipolar sources strategically distributed around the two arrays at a central distance of 3 mm from the nearest contacts on the HD-sEEG, as shown in Figure 5.

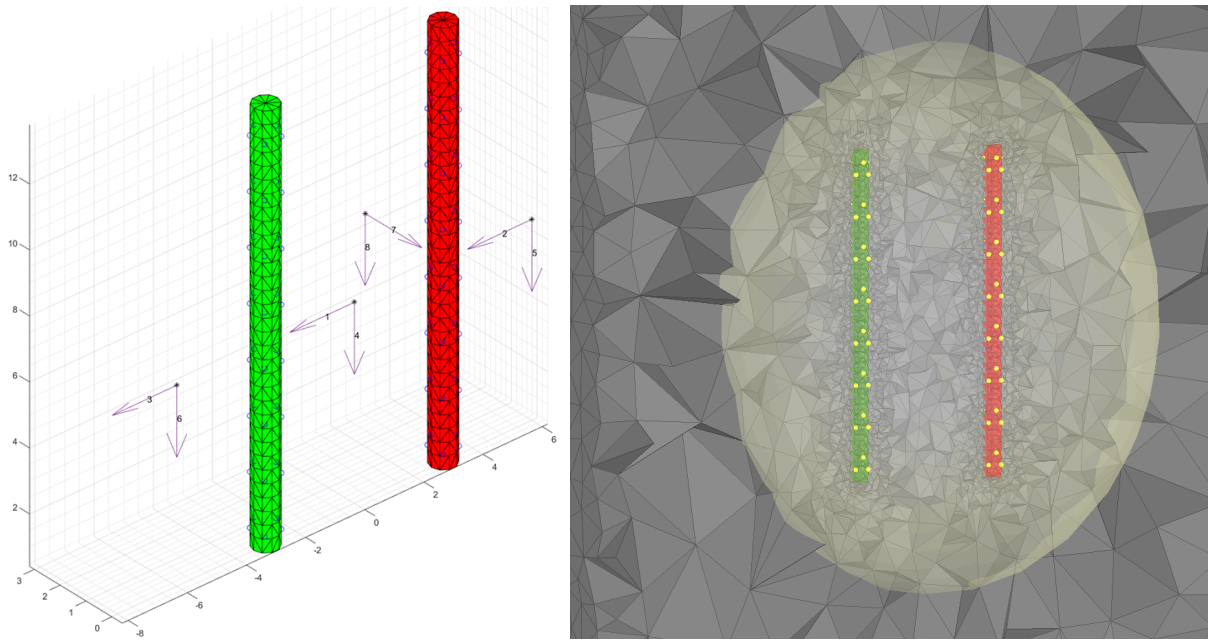

**Figure 5:** *Left*: Spatial distribution of eight simulated dipoles (purple arrows labeled with the dipole ID from 1 to 8) with different positions (3mm away from the HD-sEEG) and orientation strategically positioned around two HD-sEEG electrode arrays (red and green cylinders). *Right*: Finite Element Method (FEM) mesh of the two HD-sEEG arrays embedded within the volume conductor. The refined mesh is visible around the region of interest (ROI) highlighted by the ellipsoid, with electrode contacts plotted in yellow on the surfaces of the green and red HD-sEEG arrays.

### i- FEM Model and Electrode Configurations

Figure 5 depicts the FEM model with the two HD-sEEG arrays (shown in red and green) embedded within the brain volume in gray. The arrays contain densely spaced contacts, enabling high-resolution sampling of neural potentials. To benchmark performance, we simulated two commonly used sEEG electrode configurations by averaging signals across HD-sEEG contacts to approximate the following:

**Standard sEEG (2 mm spacing):** Signals from HD-sEEG contacts within approximately 2 mm patch were averaged to simulate a conventional clinical sEEG array (cylinder of 2mm).

**Mini-sEEG (0.5 mm spacing):** Signals from HD-sEEG contacts within approximately 0.5 mm were averaged to represent an intermediate-resolution electrode array between conventional sEEG and the dense HD-sEEG array.

Figure 6 displays the three electrode configurations side by side, showing contact arrangements and simulated coverage.

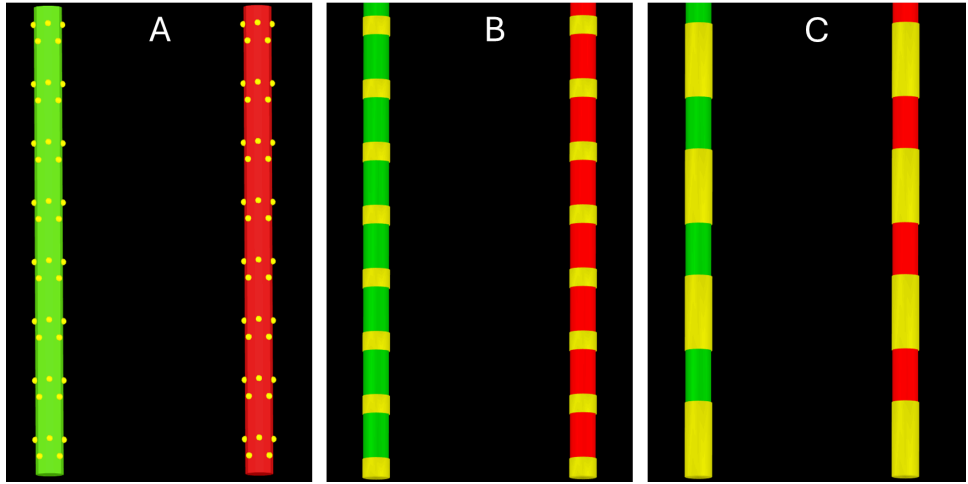

**Figure 6:** Simulated electrode configurations derived from the directional scalable (HD-sEEG) arrays. From left to right: (A) Full HD-sEEG arrays showing dense recording with 64 contacts (yellow) on the green and red shafts. (B) Mini-sEEG configuration simulated by averaging the HD-sEEG contacts, simulating 8 contacts on each shaft of a 0.5mm cylindrical contact, representing an intermediate-resolution electrode array between standard sEEG and dense HD-sEEG arrays. (C) Standard sEEG configuration simulated by averaging HD-sEEG contacts representing a 2mm conventional clinical sEEG array.

### ii- LF Computation and Dipole Localization Analysis

Each dipole was simulated statically using LFs precomputed with ANSYS. These LFs were used to generate electric potentials measurable at the HD-sEEG contacts. To mimic realistic recording conditions, variable Gaussian noise was added to create time series data. The resulting simulated signals served as input to the Brainstorm software platform for source localization analysis.

For each electrode model, the corresponding voltage signals were obtained by averaging the relevant HD-sEEG contact signals to simulate both the standard sEEG and mini-sEEG configurations. Specifically, signals from HD-sEEG contacts within approximately 2 mm distance were averaged for the standard sEEG model, while signals from contacts within about 0.5 mm apart were averaged for the mini-sEEG model.

The FEM forward computations began by constructing the geometric model of the two HD-sEEG arrays within the volume conductor (Figure 5), followed by mesh refinement around the electrode regions to improve spatial accuracy. LF were calculated separately for each electrode model: the full HD-sEEG array, mini-sEEG, and standard sEEG. For the mini-sEEG and standard sEEG models, the electrode contacts were represented as point sources, and the physical shape and conductivity of the electrode shafts were not included in the FEM modeling. Source localization analyses were then performed independently for each model using the corresponding computed LFs.

LFs were computed using the Brainstorm/DUNEuro FEM pipeline with standard parameters described previously, within a regular source grid spaced by 0.125mm (simulating around 800k sources). Finally, minimum norm estimation combined with a dipole scanning approach was applied in Brainstorm to reconstruct the positions of the simulated dipoles.

## 3. Results

We performed the simulation on all the previous models, and the obtained results are shown in the following sections. All computations were performed on a Dell Precision laptop (Intel i7 CPU, 64 GB RAM). The Brainstorm/DUNEuro pipeline made use of MATLAB scripts to import the custom electrode geometry since the Brainstorm GUI version used does not yet support user-defined sEEG/HD-sEEG probe geometry at the time of the publication of this report. ANSYS simulations were set up via MATLAB/Python scripts.

### 3.1. Validation with Spherical Head Models

To evaluate the accuracy of the numerical methods (Ansys and Duneuro), we compared their outputs against the analytical solutions for both single- and multi-layer spherical head models. The following figure shows the surface potential at the 200 electrodes.

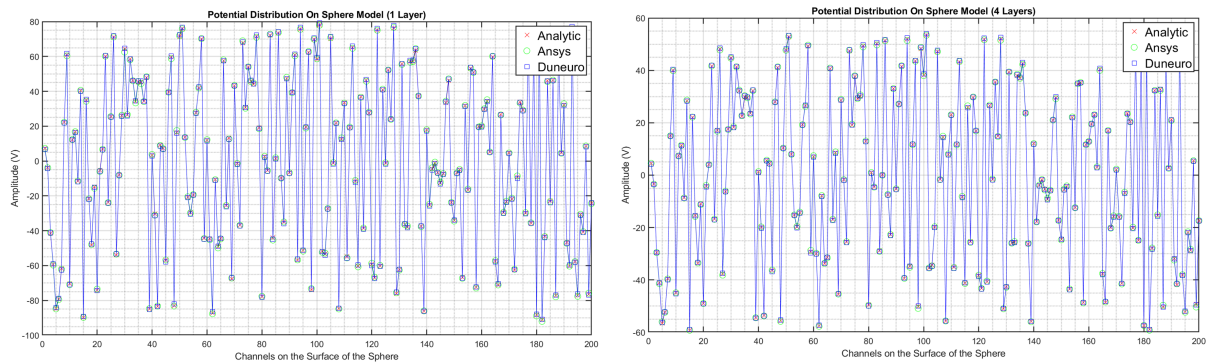

**Figure 7.** Comparison of surface potential distributions computed with Ansys and DUNEuro against the analytical solution on spherical head models. The left plot shows results for a single-layer sphere, while the right plot displays results for a four-layer sphere model. Potentials are sampled at 200 electrode positions uniformly distributed on the sphere surface generated from a random source within the inner layer. Both numerical methods (Ansys: green circles, DUNEuro: blue squares) show excellent agreement with the analytical solution (red crosses), validating their accuracy for both simplified and multilayer head geometries.

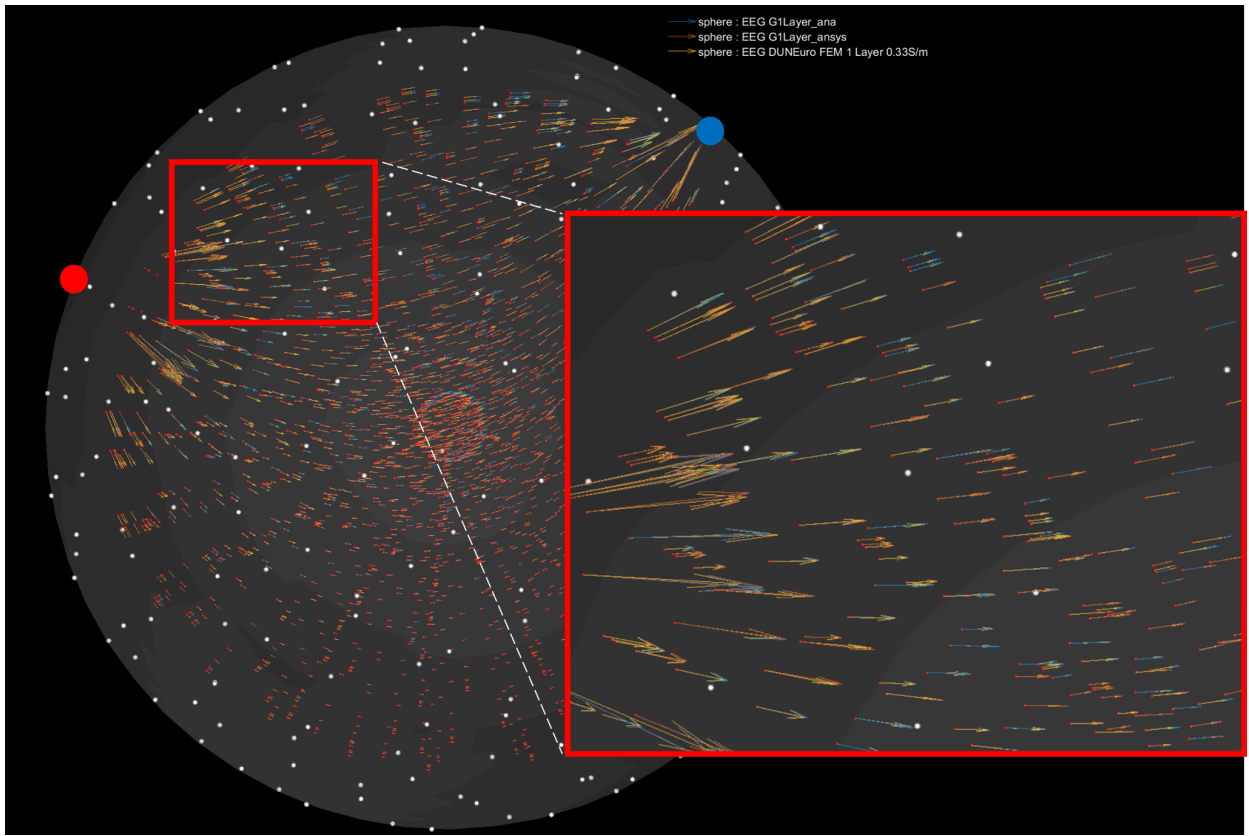

**Figure 8.** Visualization of LF vector fields computed using analytical, ANSYS FEM, and DUNEuro FEM methods in a single-layer spherical head model. The vector field illustrates the LF resulting from a bipolar electrode pair placed on the surface of the sphere (highlighted in red as anode and blue cathode). Vectors represent the spatial sensitivity of this electrode configuration to current sources within the volume. Results from the analytical solution (blue vectors), ANSYS (red vectors), and DUNEuro (yellow vectors) are shown for comparison. The inset provides a zoomed view, highlighting the close agreement in vector orientation and magnitude among the three methods, validating the numerical approaches for LF estimation. The white dots represent the location of the 200 electrodes on the surfaces, while the red dots represent the location of the sources within the inner layer.

Both FEM solvers, **Brainstorm with DUNEuro** and **ANSYS**, demonstrated excellent agreement with analytical solutions across all source configurations in the spherical head models. Surface potential distributions (Figure 7) visually matched the analytical reference, with no discernible discrepancies. In addition, the LF vector field for a bipolar electrode configuration (Figure 8) showed near-identical orientation, magnitude, and spatial distribution across the analytical, ANSYS, and DUNEuro solutions. The inset highlights this agreement at a finer scale, confirming that all methods accurately capture the directional structure and spatial gradients of the LF.

This consistency across methods reflects the reliability of FEM-based solvers in modeling electric fields, particularly in regions with strong field curvature or near active electrodes. Even subtle features, such as field divergence and focality, are faithfully reproduced, supporting the use of these numerical tools for high-fidelity modeling of electrode sensitivity and source localization in electrophysiology.

For quantitative validation, we computed three standard accuracy metrics: Relative Difference Measure (RDM), Magnitude Error (MAG), and Correlation Coefficient (CC). All solvers achieved an RDM lower than 5% indicating minimal spatial deviation in the potential distribution; MAG ≈ 1, confirming accurate amplitude scaling; and correlation CC ~98%, demonstrating near-perfect correlation between numerical and analytical solutions. These results confirm that both FEM-based methods can reliably reproduce known analytical forward solutions under idealized spherical conditions, validating their implementation for high-resolution source modeling and LF computation. Small discrepancies were observed if the meshes differed, so we ensured a sufficiently fine mesh and identical material properties to minimize interpolation error, and we ensured local mesh refinement for the next section.

### 3.2. Results for the High-Resolution Directional Electrode Model

Both Brainstorm/DUNEuro and ANSYS successfully computed the 128-channel LF of the HD-sEEG electrode model. Despite the complex geometry and high number of channels, the solutions remained consistent. The Pearson's correlation coefficient (Pearson's r) between the Brainstorm and ANSYS LF matrices was >95%, and the overall RDM was <5%. This slight discrepancy is likely due to differences in how the two solvers handle the FEM methods implementation, numerical integration of dipole sources, interpolation of the results, and the mesh resolution. When focusing on individual LFs vector (the sensitivity map of each pair of contacts), we observed that Brainstorm and ANSYS produced indistinguishable spatial patterns. For instance, Figure 10 shows LF values for the high-density HD-sEEG array. The vector fields show a characteristic dipole-like field distorted by the presence of the insulating cylinder: higher potentials on the side of the electrode facing the contact, and a "substrate shielding" of lower potential directly behind the electrode. The metrics for this contact's LF were exemplary: RDM = 5.1%, lnMAG = 0.02, CC = 98% between the two solvers. Other contacts yielded similar agreement. These results demonstrate that the open-source FEM approach can replicate the high-resolution LF calculations of the commercial software, even for a complex model such as the HD-sEEG electrodes.

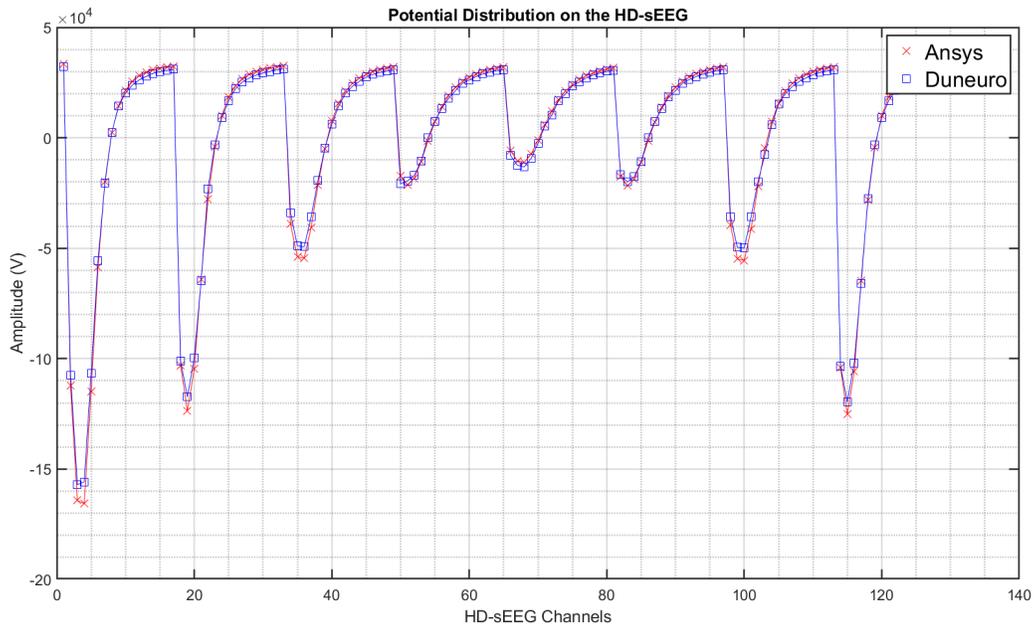

**Figure 9.** Surface potential distribution comparison across the 128 channels of the HD-sEEG electrode array computed using Brainstorm/DUNEuro (clue square) and ANSYS (red crosses). The close alignment of data points reflects high correlation (Pearson's r > 95%) and low relative difference metric (RDM < 5%), indicating excellent agreement between the open-source and commercial FEM solvers for complex high-resolution LF modeling.

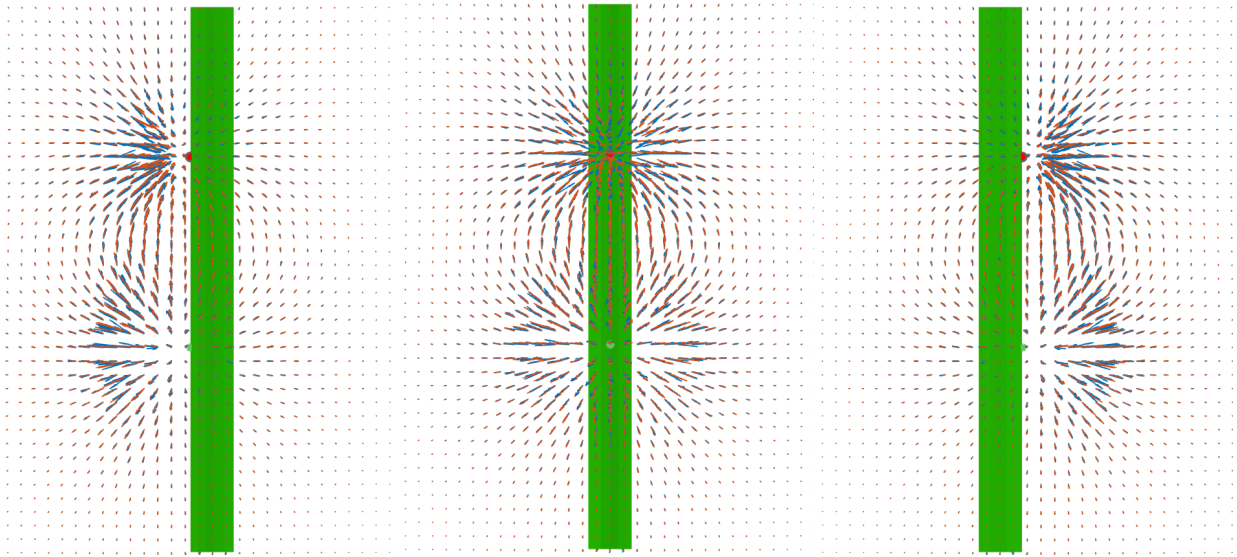

**Figure 10.** LF vector maps for the high-resolution HD-sEEG electrode model are shown from three orthogonal views. Blue arrows represent the LF vectors computed using the Brainstorm/DUNEuro FEM solver, while red arrows correspond to those computed with ANSYS. The LFs are computed specifically between two electrodes, visually highlighted in green and red. The green cylinder indicates the insulating shaft of the electrode. Both solvers produce

highly consistent spatial patterns and vector orientations, accurately capturing the characteristic dipole-like field distribution distorted by the electrode geometry..

The comparison between Brainstorm/DUNEuro and ANSYS LF vectors demonstrates excellent agreement across all three views. Vector directions and magnitudes closely overlap, indicating that both FEM solvers provide near-identical results despite differences in implementation details. This consistency validates the open-source DUNEuro solver as a reliable alternative to the commercial ANSYS software for modeling complex electrode geometries like the HD-sEEG. The high fidelity in spatial patterns confirms the robustness of both approaches for precise electrophysiological source modeling.

### i- Spatial Diversity and Directional Dependence in LF Patterns

The LF distributions for the HD-sEEG electrode model exhibit significant spatial diversity, primarily influenced by the presence of the insulating electrode shafts. Unlike an idealized isotropic source sensitivity, the physical structure of the shafts restricts current flow in certain directions, resulting in directional dependence in the LF patterns. This effect manifests as a "shadow" region of reduced sensitivity directly behind the shaft, where the insulating material blocks electrical currents. Consequently, the LF vectors show a dipole-like configuration that is distorted relative to the ideal case, with higher potentials and sensitivities oriented outward from the contact surface opposite the shaft.

This directional dependence has important implications for source localization accuracy, as it affects the spatial weighting of signals detected by each electrode contact. Accurate modeling of this spatial heterogeneity, as achieved by both Brainstorm/DUNEuro and ANSYS FEM solvers, is therefore critical for high-resolution electrophysiological mapping and interpretation in clinical and research applications.

### ii -Evaluating LF Uniformity with Homogeneous Shaft Conductivity

Building on the previous section discussing the directional dependence of the LF due to the insulating properties of the electrode shafts, this new analysis explores the impact of assigning the shaft material the same conductivity as the surrounding medium. The resulting LF isosurfaces, shown in **Figure 11**, illustrate a markedly more isotropic and symmetric spatial sensitivity pattern compared to the original model with insulating shafts.

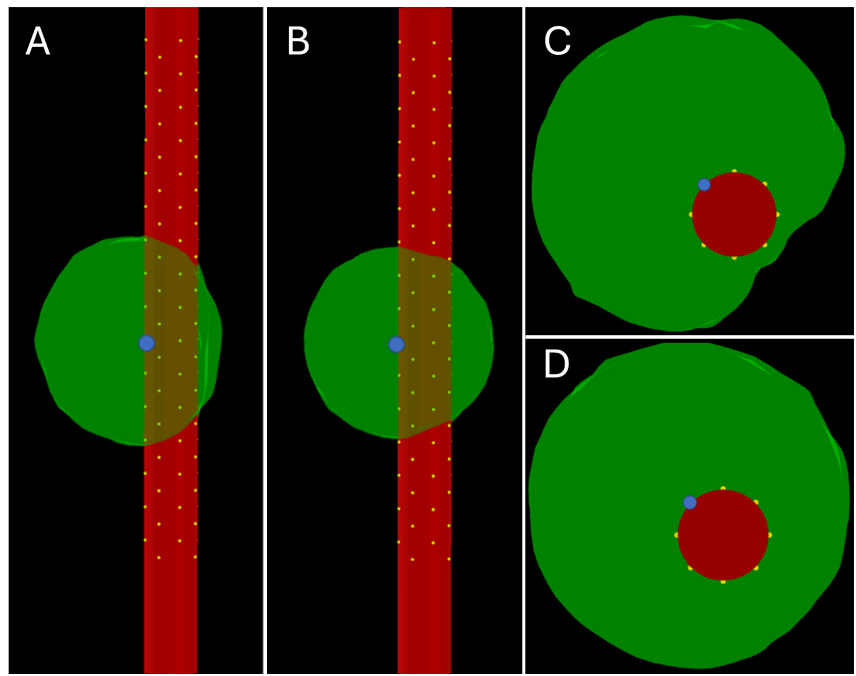

**Figure 11.** LF isosurfaces for the high-resolution HD-sEEG electrode model comparing the effects of shaft conductivity, and demonstrated the "substrate shielding" effect. Panels A and B show front views of the electrode shaft (red cylinder) and LF (green isosurface) around the selected electrode (blue sphere) with regard to distant reference for the insulating shaft model (A) and the homogeneous conductivity shaft model (B). Panels C and D show corresponding top-down views for the insulating shaft (C) and homogeneous shaft conductivity (D) conditions. Assigning the shaft the same conductivity as the surrounding tissue produces a more isotropic and symmetric LF distribution, reducing the spatial heterogeneity and directional dependence observed with the insulating shaft.

Panels **A** and **B** of Figure 11 show front views of the electrode shaft (red cylinder) and LF (green isosurface) around the selected electrode (blue sphere), contrasting the insulating shaft model (A) with the homogeneous conductivity shaft model (B). Corresponding top-down views for these conditions are depicted in panels **C** and **D**, respectively. Assigning the shaft the same conductivity as the surrounding tissue produces a more isotropic and symmetric LF distribution, substantially reducing the spatial heterogeneity and directional dependence observed in the insulating shaft model.

This outcome confirms that the insulating shaft geometry is a primary factor inducing spatial heterogeneity and directional dependence in the LF distribution. When the shaft conductivity is homogenized with the surrounding tissue, the "substrate shielding" effect and directional bias previously observed are substantially reduced, resulting in smoother and more uniform LF profiles.

These results underscore the importance of accurately modeling electrode shaft conductivity to capture realistic spatial diversity in sensitivity. Such precision is crucial for improving source localization fidelity in applications relying on directional electrode arrays like the HD-sEEG.

### 3.3. Dipole Localization Accuracy Across Electrode Configurations

We quantitatively evaluated the localization accuracy of simulated dipolar sources using three electrode configurations derived from the HD-sEEG arrays: the full HD-sEEG array, the mini-sEEG array, and the standard sEEG array, as explained in section 2.5. Localization performance was measured by computing the Euclidean distance between the true dipole positions and the reconstructed dipole locations for each of the eight simulated sources (see sections 2.5).

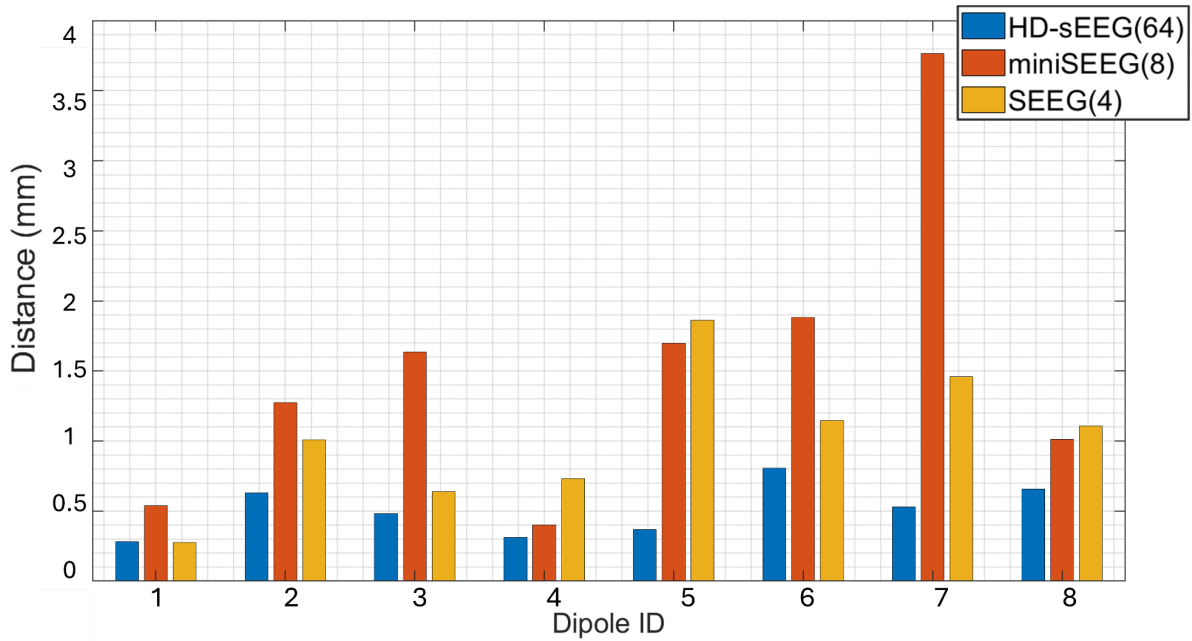

**Figure 12** Euclidean distances between true dipole positions and their reconstructed locations for eight simulated dipoles across three electrode configurations: full HD-sEEG array (blue bars, 64 contacts), mini-sEEG (orange bars, 8 contacts), and standard sEEG (yellow bars, 4 contacts). The HD-sEEG array consistently shows the lowest localization errors, demonstrating improved spatial precision compared to the sparser mini-sEEG and standard sEEG configurations. Error distances for each dipole highlight the benefits of high-density electrode arrays in accurately resolving neural sources.

As illustrated in the bar plot, the full HD-sEEG array (blue bars) consistently achieved the lowest localization errors across all dipoles, with distances typically under 1 mm. This underscores the advantage of high-density contacts for precise source estimation. The standard sEEG configuration (yellow bars), which reflects conventional clinical electrode spacing, showed moderately higher localization errors, generally ranging between 0.5 and 2 mm. The mini-sEEG configuration (orange bars), representing an intermediate contact density, exhibited the greatest variability and overall higher errors, with some dipoles exceeding 3.5 mm in localization error.

Beyond positional accuracy, we also assessed the reconstructed dipole amplitudes and orientations. These results paralleled the distance findings: the HD-sEEG array yielded more accurate amplitude and orientation estimates compared to both mini-sEEG and standard sEEG arrays. This further validates the superior spatial resolution and source discrimination capabilities afforded by the high-density HD-sEEG design.

Together, these findings highlight the critical influence of electrode contact density and spatial sampling on source localization performance. The dense HD-sEEG arrays enable finer spatial resolution, resulting in improved reconstruction of neural sources relative to sparser configurations. The increased errors observed in the mini-sEEG condition likely reflect the trade-off between contact spacing and the ability to resolve closely spaced sources, emphasizing the value of high-resolution HD-sEEG arrays for advanced electrophysiological mapping.

## 4. Discussion

This study provides a comprehensive evaluation of FEM-based LF modeling for novel high-density directional depth electrodes, demonstrating that open-source tools like Brainstorm/DUNEuro achieve accuracy comparable to commercial software such as ANSYS. This validation supports broader adoption of accessible, integrated pipelines within the neuroscience community.

### 4.1. Solver Concordance and Open-Source Benefits

The close agreement between Brainstorm/DUNEuro and ANSYS across both simple and complex models confirms the correctness of the physical modeling in both tools. Brainstorm/DUNEuro's seamless integration with neuroimaging workflows, covering MRI segmentation, meshing, and LF computation and other tools for source analysis and modeling, offers significant advantages over general-purpose FEM commercial software, including reproducibility and cost-free access. While custom code was necessary to represent high-resolution electrode geometries, future Brainstorm updates may incorporate these capabilities natively.

### 4.2. Directional Electrode Effects and Clinical Implications

A key finding is the significant impact of modeling the electrode's physical structure on LF distributions. The insulating shaft of directional electrodes causes **substrate shielding**, amplifying signals on the contact-facing side, and creating directional sensitivity. This spatial diversity affects the interpretation of LFP recordings, as signal amplitude reflects both source proximity and directional alignment. Incorporating these detailed LFs into source localization can improve spatial accuracy, benefiting applications such as epileptic focus localization and adaptive deep brain stimulation (DBS). Our results quantitatively support prior findings on the enhanced spatial information provided by directional electrodes.

### 4.3. Technical Challenges

**i- Mesh Resolution:** Accurate modeling of fine electrode geometries HD-sEEG requires very dense tetrahedral meshes, which substantially increases memory and runtime. Simply refining the entire volume is inefficient. Adaptive or multi-resolution meshing, refining only in the vicinity of electrodes and conductivity discontinuities while coarsening elsewhere, offers a practical compromise, but robust automatic criteria for refinement still need optimization.

**ii- Tissue Anisotropy:** We assumed isotropic conductivities for simplicity, omitting white-matter anisotropy known to influence current flow and the spatial distribution of LFs. Incorporating diffusion MRI–derived tensors would improve physiological realism and may alter sensitivity profiles, particularly for deep sources [4,36,37]. Future work will evaluate how anisotropy interacts with electrode geometry and whether simplified reduced-tensor models can capture most of the benefit at lower computational cost.

**iii- Reciprocity, Conductivity and Solver parameters:** Consistent application of reciprocity principles, boundary conditions, and reference scaling across numerical solvers is essential to avoid systematic offsets. Uncertainties in tissue conductivities primarily affect absolute voltage amplitudes rather than relative lead-field topographies, but propagate into inverse solutions when comparing modalities. In addition, we observed inconsistencies in solver post-processing, including smoothing and interpolation of the numerical solution, which are not documented in ANSYS and may introduce further discrepancies between platforms. Establishing standardized conductivity priors, automated unit checks, explicit control of post-processing operations, and cross-solver validation benchmarks will help mitigate these sources of error.

### 4.4. Future Directions

**i- Multiscale Modeling:** Coupling FEM LFs with cellular-level neural simulations enables detailed evaluation of electrode sensitivity to microcircuit activity.

**ii- Computational Efficiency:** Development of adaptive meshing, model order reduction, and machine learning surrogates can accelerate FEM computations and enable rapid exploration of electrode designs.

**iii- Experimental Validation:** Planned saline tank experiments with controlled dipole sources will test model predictions and reveal real-world effects such as contact impedance.

**iv- Clinical Translation:** Applying this pipeline to patient-specific data can enhance source localization accuracy and improve configuration of sensing-enabled DBS devices.

### 4.5. Limitations

The directional electrode simulations employed a homogeneous brain model lacking anatomical detail and tissue heterogeneity. Embedding electrodes in realistic head models is necessary to evaluate anatomical influences on LF patterns. Contact–tissue interface properties and capacitive effects were not considered, but are less relevant for low-frequency signals, as in our study.

## 5. Conclusion

We have presented a comprehensive FEM modeling study of a high-resolution, directional depth electrode, rigorously comparing results from a commercial solver (ANSYS) and the open-source Brainstorm/DUNEuro pipeline. This work delivers two major contributions: (1) a thorough validation of the Brainstorm/DUNEuro framework for accurate LF computation, supporting its broader adoption as a cost-effective, integrated alternative to commercial software, and (2) novel insights into how electrode design, particularly the insulating substrate and geometry, shapes directional recording sensitivity and spatial selectivity.

By providing validated models and methods, we enable researchers and clinicians to simulate current and novel electrode designs with improved accuracy. Our findings emphasize the importance of detailed forward modeling for precise source localization and adaptive device development.

Advanced FEM modeling reveals how electrode design influences recorded signals, supporting optimization of spatial resolution and directional selectivity. These benefits include applications such as epilepsy monitoring, deep brain stimulation, and future high-density brain-computer interfaces.

The continued advancement of neural hardware alongside open-source modeling tools promises to accelerate understanding and modulation of brain function at finer spatial scales.

## 6. Data and code availability statement

The data and code supporting the findings of this study are available from the corresponding author upon reasonable request.

## 7. Acknowledgments

Research reported in this publication was supported by the National Institute of Biomedical Imaging and Bioengineering (NIBIB) of the National Institutes of Health (NIH) under award numbers R01EB026299 and RF1NS133972.

## 8. Authorship contributions

**T.M.** led the conceptualization, design, development, simulation, and evaluation of the computational framework. **J.W.** and **C.W.** contributed to the conceptual design, development, simulation, evaluation, and interpretation of the results. **Y.V.** contributed to the development protocol and the interpretation. **R.S.** and **R.C.** contributed to the model development and implementation. **A.J.** assisted with the development and evaluation of the computational framework. **R.L.** provided scientific supervision and oversight throughout the project. **J.S.** and **J.M.** contributed to the conceptualization, design, development, and evaluation of the computational framework as well as scientific supervision, interpretations of the results, and oversight throughout the project. All authors reviewed and approved the final version of the manuscript.